\documentclass[a4paper,11pt]{article}
\usepackage{pos}

\title{Comments on the Emergence of 4D Topological  Amplitudes in M-Theory}
\author[a]{Manuel Artime}
\author*[a]{Ralph Blumenhagen}
\author[a]{Aleksandar Gligovic}
\author[a,b]{Panagiotis Leivadaros}

\affiliation[a]{Max-Planck-Institut f\"ur Physik,\\
   Boltzmannstrasse 8,  85748 Garching, Germany }
\affiliation[b]{Fakult{\"a}t f{\"u}r Physik, Ludwig-Maximilians-Universit{\"a}t M\"unchen, \\ 
  Theresienstr.~37, 80333 M\"unchen, Germany}

 \emailAdd{artime@mpp.mpg.de}
 \emailAdd{blumenha@mpp.mpg.de}
 \emailAdd{aglig@mpp.mpg.de}
\emailAdd{p.leivadaros@campus.lmu.de}

\abstract{The M-theoretic Emergence Proposal claims that all of the
  terms in the low-energy action arise from quantum effects. After
  reviewing the current status of this proposal, we focus on
  four-dimensional compactifications with $N=2$ supersymmetry, where
  kinetic terms are encoded in topological string amplitudes, such as
  the prepotential $\mathcal{F}_0$. Evidence for the emergence of such
  terms was provided recently, where in
   particular it was shown that the classical cubic term in ${\cal F}_0$ can be obtained by integrating out the light towers of states in the M-theory limit, using a novel regularization of the infinite sum over Gopakumar/Vafa invariants. We address two issues that were left open. First, we show that the regularization can be equivalently performed in complex structure moduli space and in Kähler moduli space. Second, we extend the proposed regularization to the linear terms in the one-loop prepotential $\mathcal{F}_1$.}

\FullConference{Corfu Summer Institute 2025 "School and Workshops on Elementary Particle Physics and Gravity" (CORFU2025)\\
 27 April - 4 May , and 24 August - 28 September, 2025\\
Corfu, Greece\\}

\newcommand{\eq}[1]{\begin{equation}
	\begin{split} #1 \end{split}
	\end{equation}}

\newcommand{\nc}{\newcommand}
\nc{\lb}{\llbracket}
\nc{\rb}{\rrbracket}
\nc{\gl}{\llbracket}
\nc{\gr}{\rrbracket}
\nc{\del}{\partial}
\nc{\fa}{\hat}
\nc{\fb}{\MakeUppercase}
\nc{\fc}{\tilde}
\nc{\myhash}{\raisebox{\depth}{\#}}

\begin{document}
\renewcommand{\hookAfterAbstract}{
    \par\bigskip\bigskip\bigskip
    Report number: MPP-2026-40
}
\maketitle

\section{Introduction}
\label{sec_intro}

One of the central aspects of the   swampland program
(see \cite{Palti:2019pca,vanBeest:2021lhn,Agmon:2022thq} for reviews)
is the realization that string theory, or more generally Quantum Gravity (QG), admits infinite distance limits in its moduli space where
infinite towers of states become exponentially light.
This is captured by the swampland distance conjecture \cite{Ooguri:2006in},
which has been analyzed and refined during recent years.
One essential refinement is the emergent string conjecture \cite{Lee:2019wij},
which asserts that the lightest of such towers
can only be of two types. These are either Kaluza-Klein (KK) modes
of a decompactifying direction or the excitation modes
of a weakly coupled, critical string, both meant in an
appropriate duality frame.

In addition to the lightest towers appearing at infinite-distance limits, it was found that
there exists a 
hierarchical pattern of all towers, namely that one can distinguish between
light and heavy towers  \cite{Blumenhagen:2023tev,Blumenhagen:2023xmk}, where the former ones are
those with a typical mass scale not larger than the QG cut-off, which is taken to be the species scale \cite{Dvali:2007hz,Dvali:2007wp}.
Similar to the structure of the weakly coupled
fundamental string, the light towers can be considered
the fundamental quantum degrees of freedom
and the heavy ones as  classical non-perturbative contributions.
From this perspective, standard perturbative string theory is just
the perturbative QG theory that arises in the small string coupling
regime, i.e. an  infinite distance limit for the dilaton.
The perturbative states are the vibration modes of the string and upon
compactification also its KK and winding modes.
The non-perturbative states are the $NS5$-brane together with the various $D$-branes of
the superstring theories, all with tensions larger than the species scale.

In an attempt to find a microscopic explanation to the swampland distance conjecture the so-called Emergence Proposal
\cite{Heidenreich:2017sim,Grimm:2018ohb,Heidenreich:2018kpg} was put forward
(for subsequent work
see e.g. \cite{Marchesano:2022axe,Castellano:2022bvr,Blumenhagen:2023yws}),
which suggests that
the kinetic terms in the low-energy effective action of
(all) QG theories are emerging quantum mechanically from integrating out 
states below a certain UV scale.
While initially less understood than some other swampland conjectures, it was subsequently realized that there exists a special limit in the QG moduli space where the conjecture seems to be realized in an even stronger sense \cite{Blumenhagen:2023tev,Blumenhagen:2023xmk,Hattab:2023moj}
(for a complementary approach see \cite{Hattab:2024thi,Hattab:2024chf,Hattab:2024ssg,vanMuiden:2026nsp}).
First, consider a compactification of type IIA string theory to $d$ dimensions on an internal space with radii $R_i$. The M-theory limit is
obtained by scaling the type IIA string coupling and the radii to
infinity in such a way that, in M-theory units, one has
\eq{
  \label{Mtheorylimit}
  R_{11}\to \lambda R_{11}\,,\qquad  
  M_*\to \lambda^{\frac{1}{d-1}} M_*\,,
  \qquad 
  M_* R_i \to 1\,,
}
where $R_{11}$ denotes the radius of the M-theory circle and $M_*$ the
eleven-dimensional Planck scale. In this limit one takes the strong
coupling regime of type IIA string theory, in which a new eleventh
dimension opens up. At the same time, the size of the internal space
in units of the eleven-dimensional Planck scale, as well as the
$d$-dimensional Planck scale, are kept finite. This limit is called
the M-theory limit \cite{Blumenhagen:2023xmk}  and is a
decompactification to $d+1$ dimensions where the species scale is the
higher dimensional Planck scale. Second, it was proposed that in this
limit the full low-energy effective action arises from integrating out
the light towers of states, namely those whose typical mass scales do
not exceed the species scale. In other words, all interactions are
generated through quantum effects, without any classical contribution
(see \cite{Blumenhagen:2024lmo} for a review).

The central challenge in providing evidence for this proposal is that
we do not understand the full quantization of M-theory yet.
Guided by the BFSS matrix model \cite{Banks:1996vh} (see \cite{Taylor:2001vb} for a review), one realizes that this form of the Emergence  Proposal is consistent with it. There, the fundamental degrees of freedom are matrices and a large scale commuting space-time can arise by solving the classical equations
of motion, in particular by choosing
diagonal matrices.
However, interactions like gravity are known to 
be obtained from a quantum effect, namely from
integrating out the fluctuations of matrices around
the classical background.

However, for more involved compact backgrounds, a similar matrix model computation is not feasible with current techniques. Nonetheless, these  difficulties can be bypassed by considering certain couplings in the effective action that are protected by
supersymmetry and only receive
contributions from $1/2$-BPS states. The latter are well
understood and  admit the usual
geometric interpretation we are used to from string theory.
Indeed, all current evidence for the M-theoretic Emergence Proposal are  based on the evaluation of such $1/2$-BPS saturated couplings. Explicit checks were carried out by deriving these couplings from a one-loop Schwinger integral, where only the light, perturbative towers of particle-like states with masses below the species scale, namely the 11D Planck scale, are integrated out.

Let us briefly recall which consistency checks have been performed:
\begin{itemize}
\item{\textbf{$\mathbf{R^4}$-term in supergravity theories with 32 supercharges:}
    Arising
from toroidal compactifications of type IIA/M-theory, the analysis
builds on the pioneering work of Green-Gutperle-Vanhove \cite{Green:1997as},
in which the couplings in ten and nine dimensions were obtained by integrating out
the KK spectrum along the eleventh direction, i.e. bound states of
$D0$-branes.
In \cite{Blumenhagen:2024ydy} tools were developed 
to regularize the appearing Schwinger integrals and shown
to be consistent  with string theory results.
This was then applied to the evaluation of the $R^4$ Schwinger integral, where the rest of the light towers of states were integrated out. For dimensions $d\ge 7$ integrating out the full set of light towers leads to a perfect match with the
exact amplitude.
This analysis will be  further
refined in \cite{Artime:2026new}, where the various infinite distance
limits as classified by the taxonomy rules \cite{Etheredge:2024tok}
will be  analyzed in a systematic manner.
Importantly, only the M-theory limit 
from  eq.\eqref{Mtheorylimit} remains as the limit where the full $R^4$-term arises
from integrating out the light modes.}

\item{$\mathbf{F^4}$\textbf{-term in six-dimensional supergravity with 16 supercharges:} This gauge coupling is obtained by
compactifying type IIA on a $K3$ surface or the dual heterotic string on
a $T^4$. Building on \cite{Kiritsis:2000zi} and exploiting the duality at the special orbifold point $T^4/\mathbb Z_2$ in the $K3$ moduli space,
\cite{Artime:2025egu} focuses on the $F^4$-coupling
for a linear combination of 16 $U(1)$ gauge fields, each arising from
the dimensional reduction of the Ramond - Ramond (RR) 3-form $C_3$ along a two-sphere. With
a careful case-by-case study of the worldsheet instanton corrections
on the heterotic side, it was possible to show that $D4$-brane
contributions are indeed redundant, in the sense that they lead to
mutually canceling terms, as their mass scale is above the species scale.}

\item{\textbf{The topological string couplings ${\cal F}_0$ and ${\cal F}_1$ in four-dimensional $N = 2$
supergravity (8 supercharges):} These are related to F-terms in the
effective action of type IIA compactified on a Calabi-Yau (CY) threefold. Note
that ${\cal F}_0$ contains information about the second order supergravity
action, namely about the gauge couplings and kinetic terms of the
vector multiplet fields. Integrating out the light towers
is precisely what has been done in the seminal work
of Gopakumar/Vafa \cite{Gopakumar:1998ii,Gopakumar:1998jq}, where the authors expressed the topological string
amplitudes ${\cal F}_g$ for genus $g \ge  0$ in terms of a Schwinger integral, integrating
out particle-like bound states of $D0$- and $D2$-branes.
Although originally only obtaining the worldsheet instanton contributions, it was shown in \cite{Blumenhagen:2023tev} that for the  non-compact resolved conifold also the classical
polynomial contributions could be obtained by an appropriate
regularization of the UV divergences. These results  were
the building blocks to generalize the computation
to compact CY manifolds. In this case, the main challenge
is a final regularization of the infinite sum
over the so-called Gopakumar/Vafa (GV) invariants, which generically grow
exponentially. This problem was approached in \cite{Blumenhagen:2025zgf},
where we provided a concrete recipe of how such
a regularization procedure can be implemented.
A key aspect of the regularization is that it involved taking a limit to certain singularities
in the K\"ahler moduli space. However, for technical reasons, this limit was actually
taken in the mirror dual complex structure moduli space.}
\end{itemize}

In this article we focus the attention
on this last setup of topological couplings  ${\cal F}_0$ and ${\cal F}_1$ in four-dimensional $N = 2$
supergravity. We first summarize the main idea of \cite{Blumenhagen:2025zgf}, which provided a regularization of the sum over all GV invariants for ${\cal F}_0$. Instead of revisiting in detail the computation presented there, this proceedings article focuses on two specific issues that emerged during discussions of this work. On the one hand we reconsider the regularization of ${\cal F}_0$, originally performed in the mirror complex structure moduli space, and confirm whether it yields the same result as taking the corresponding limit in the original Kähler moduli space. On the other hand, it is known that also ${\cal F}_1$ receives polynomial, in fact linear, contributions so a natural question is whether our regularization procedure can also be extended to this case. In both cases we approach these problems for the case of the Quintic.

\section{Emergence of 4D \texorpdfstring{$N=2$}{N=2} topological couplings}

It is known that the topological amplitudes ${\cal F}_g$ in type IIA compactifications on  CY threefolds are $1/2$-BPS saturated. This yields $N=2$ supersymmetry in four dimensions, and  the ${\cal F}_g$ only depend on the K\"ahler moduli. Since this is a well studied class of models, let us only very briefly recall some relevant facts.

\subsection{Preliminaries}

The corresponding vector-multiplet moduli space is spanned by real
scalars $\tau_i$ which, together with the Kalb-Ramond axions $b_i$,
define the complex Kähler moduli $t_i = b_i + i\tau_i$. 
The kinetic terms and gauge couplings of the vector-multiplets are
determined by a holomorphic prepotential $\mathcal{F}_0(t)$. Due to
supersymmetry, this is tree-level exact in type IIA and
for a CY with a single K\"ahler modulus reads
\begin{equation}
\label{prepot}
\mathcal{F}_0(t) =  \frac{(2\pi i)^3}{g_s^2} \Big[
\frac{1}{6}\kappa_{ttt} \, t^3 - \frac{\zeta(3)}{2(2\pi i)^3}
\ \chi + \frac{1}{(2\pi i)^3}\sum_{\beta \in H_2(X,\mathbb{Z})}
\alpha_0^{\beta} \, {\rm Li}_3(e^{2\pi i\beta \cdot t}) \Big] \,,
\end{equation}
where $\kappa_{ttt}$ denotes the triple intersection
number of the CY $X$,
$\chi$ its Euler characteristic and $\alpha_0^{\beta}$ are genus
zero GV invariants \cite{Gopakumar:1998ii,Gopakumar:1998jq}\footnote{
 Note that mirror symmetry fixes the a priori ambiguous quadratic and
 linear terms in the prepotential,
 which we are ignoring here.}. 
In addition there are higher genus contributions ${\cal F}_g$
to the prepotential, which are  the coefficients of  higher
derivative corrections.
Similarly to the triple intersection number in
$\mathcal{F}_0$, the genus-one prepotential $\mathcal{F}_1$ contains a
classical term at weak coupling, which for  a one-parameter CY threefold reads
\begin{equation}
  \label{F1}
    \mathcal{F}_1(t)={2\pi i}\ \left(\frac{c_2}{24}t-\frac{1}{2\pi i}\sum_{\beta\in H_2(X,\mathbb{Z})}\left(\frac{1}{12}\alpha_0^\beta+\alpha_1^\beta\right){\rm Li}_1(e^{2\pi i\beta \cdot t})\right)\,,
\end{equation}
where $c_2$ is the second Chern class of the manifold and $\alpha_1^\beta$ are the genus one
GV invariants \cite{Gopakumar:1998ii,Gopakumar:1998jq}.

The M-theoretic Emergence Proposal asserts that the full expressions can be reproduced in the M-theory limit \eqref{Mtheorylimit} from a single Schwinger integral. In this limit, the M-theory scaling corresponds to rescaling all type IIA K\"ahler moduli as $\tau_i \rightarrow \lambda \, \tau_i$ and the string coupling as $g_s \rightarrow \lambda^{3/2} g_s$, with $\lambda\rightarrow\infty$, while keeping the four-dimensional Planck scale fixed. In this regime, the $D0$-branes become the lightest states and determine the species scale $\Lambda_{\rm sp} \sim M_{\rm pl}/\lambda^{1/2}$, which lies at the threshold of the mass of $D2$-branes wrapping 2-cycles of the CY manifold. Expressed in M-theory units, the co-scaled limit keeps the radii of the CY manifold fixed when measured in units of the eleven-dimensional Planck scale $M_*$, while the M-theory circle radius and the Planck scale scale as $r_{11} \rightarrow \lambda \, r_{11}$ and $M_* \rightarrow M_*/\lambda^{1/2}$, respectively. In such a limit the species scale coincides with the five-dimensional Planck scale, which scales in the same way as $M_*$.

Integrating out the light towers of states with mass scale below
or at the species scale we arrive at the prescription of Gopakumar/Vafa
\cite{Gopakumar:1998ii,Gopakumar:1998jq}, i.e.\ one has to   integrate
out
wrapped $M2$-branes transverse with respect to the
M-theory circle but with KK-momentum along it.
The corresponding Schwinger integral for $\mathcal{F}_0$  reads
\begin{equation}
\label{F0_Schwinger}
\mathcal{F}_0(t) = \sum_{(\beta,n) \neq (0,0)} \alpha_0^{\beta}
\int_0^{\infty} \frac{ds}{s^3} \; e^{s Z_{\beta,n}} \,,
\end{equation}
where $Z_{\beta,n}= \frac{2\pi i}{g_s}(\beta \cdot t - n)$ is the
central charge of the supersymmetry algebra.
Similarly, for $\mathcal{F}_1$ Gopakumar/Vafa obtained
\begin{equation}
  \label{F1_Schwinger} 
    \mathcal{F}_1(t)= -\sum_{(\beta,n)\neq (0,0)}\left(\frac{1}{12}\alpha_0^\beta+\alpha_1^\beta\right)\int_0^\infty \frac{ds}{s}e^{sZ_{\beta,n}}\,.
\end{equation}
We recall that, after employing a minimal subtraction of the UV
divergence at $t\to 0$ and subsequently applying
$\zeta$-function regularization,
the complete  $\mathcal{F}_0$ and $\mathcal{F}_1$  for the (non-compact)
resolved conifold
have been reproduced in \cite{Blumenhagen:2023tev}. This included
also the cubic and linear term in $\mathcal{F}_0$ and $\mathcal{F}_1$,
respectively. In that case, the sum over GV invariants in \eqref{F0_Schwinger} and \eqref{F1_Schwinger} was absent. However, 
for a general compact CY, the $D2$-$D0$
bound states wrap all 2-cycles in $H_2(X,\mathbb{Z})$ and therefore
another infinite sum needs to be performed.

How such a regularization of the infinite sum
over GV invariants can be performed was analyzed
in \cite{Blumenhagen:2025zgf}, at least for the prepotential $\mathcal{F}_0$.
It is easier to evaluate its  third derivative with respect to
$t$, giving the Yukawa coupling, for which the
integral over $t$ can be performed explicitly
\begin{equation}
  \label{hannover96}
  Y_{ttt}=\frac{g_s^2}{(2\pi i)^3}\partial_t^3 \mathcal{F}_0(t) =
\sum_{\beta \neq 0} \alpha_0^{\beta}  \,\beta^3 \left(\frac{1}{2} +
  \frac{e^{2\pi i \beta t}}{1-e^{2\pi i \beta t} } \right) \,.
\end{equation}
Here the second term in the brackets is related to the
non-perturbative terms in \eqref{prepot}, i.e.~the $ {\rm
  Li}_3$-terms and the factor $1/2$ matches precisely the formal triple
intersection number for the single two-cycle of the resolved conifold
\cite{Gopakumar:1998ki}.
Note that for a compact CY, the first term still involves
the infinite sum over all GV invariants,
which was called the  {\it zero point Yukawa
  coupling} in \cite{Blumenhagen:2025zgf} and was denoted as $Y_{ttt}^{(0)}$.
This object is the focus of our interest, as the Emergence
Proposal implies that computing this quantity in the strongly coupled M-theory limit should give the triple intersection number of the CY.
To appreciate the task,
recall that the exact determination of $Y^{(0)}_{ttt}$
in a fully fledged microscopic description of M-theory
would (likely) lead  to  a finite result right away.
Since such a theory is lacking at present,
the best one can try for is to find  a properly working regularization
scheme.

In the following we do not repeat the detailed derivation leading to this proposal and instead provide a heuristic motivation. For the complete computation we refer the reader to the original work \cite{Blumenhagen:2025zgf}. Implementing the zeta function regularization
$\sum_{k=1}^\infty 1=\zeta(0)=-1/2$ one writes
\eq{
  \label{regula}
  \frac{1}{2}\sum_{n=1}^\infty  \alpha_0^n n^3 = -\sum_{n,k=1}^\infty  \alpha_0^n n^3 
 =-\lim_{\tau\to 0} \sum_{n,k=1}^\infty  \alpha_0^n n^3 e^{-2\pi n k
   \tau}
 = -\lim_{\tau\to 0} \sum_{n=1}^\infty  \alpha_0^n n^3 \frac{q^n}{1-q^n}
 }
 with $q=\exp(-2\pi \tau)$.
Here, one recognizes the last term as the  instanton terms in
\eqref{hannover96}. To evaluate that sum one needs to determine the GV invariants.  
Determining these invariants by direct counting is computationally out of reach in general; however,  their values are known from mirror symmetry. Hence, setting for ease of notation  $(2\pi i)^3/g_s^2=1$, we can express the last term in \eqref{regula}
as
\eq{
  \label{regulb}
   \frac{1}{2}\sum_{n=1}^\infty  \alpha_0^n n^3
 = -\lim_{\tau\to 0} \sum_{n}^\infty  \alpha_0^n n^3 \frac{q^n}{1-q^n}
 =-\lim_{t\to 0}  \left[(\del_t^3 \mathcal{F}_0|_{\rm weak}-\kappa_{ttt})\right]
}
where we have identified  $t=i\tau$ and $\mathcal{F}_0|_{\rm weak}$
denotes the prepotential that one determines
by determining periods in the complex structure moduli and
applying the  mirror map. This formula now opens up
the possibility to define the regularization of the zero-point Yukawa
coupling by minimal subtraction of the divergent terms
in the limit $t\to0$ . Hence, we define the regularization
\begin{equation}
 \label{regyukawapre}
    Y_{ttt}^{(0)}=-\lim_{t\rightarrow 0}\left[(\del_t^3 \mathcal{F}_0|_{\rm weak}-\kappa_{ttt})-{\rm Div}\right]\,,
\end{equation}
where Div denotes all terms that diverge in the limit.

Nonetheless, there is still the issue that  it  involves taking the limit $t \rightarrow 0$.
In fact, as already observed in 
\cite{Candelas:1990rm}, 
the exponential degeneracy of the GV invariants is correlated with the appearance 
of a divergence in the prepotential, not at zero but at a finite value 
$t=t_c$ determined by the location of the conifold singularity. 
Indeed, the point $t=0$ is not even part of the quantum moduli space. 
Moreover, $t_c$ is not the minimal value of $\mathrm{Im}(t)$, 
which is instead given at the Landau--Ginzburg point. 
The conifold point is nothing else than the quantum-corrected singular point 
in the middle of the classical phase diagram of a Gauged Linear Sigma Model (GLSM).

In this sense, the limit can be interpreted as approaching zero size, corresponding to the vanishing of the Fayet--Iliopoulos (FI) parameter, $\xi \to 0$. However, since it is known (see e.g. \cite{Morrison:1994fr}) that due to quantum effects the location of the singularity changes, the most natural thing to do is to take the limit towards its quantum corrected value so that we define
\begin{equation}
 \label{regyukawa}
    Y_{ttt}^{(0)}=-\lim_{t\rightarrow t_c}\left[(\del_t^3 \mathcal{F}_0|_{\rm weak}-\kappa_{ttt})-{\rm Div}\right]\,.
  \end{equation}
A few remarks about this prescription are in order:
\begin{itemize}
\item{Since the wanted Yukawa coupling $\kappa_{ttt}$ already explicitly
    appears on the right hand side, the procedure
    works if the limit of $\del_t^3 \mathcal{F}_0|_{\rm weak}$ only
    involves divergent terms and terms that vanish in the $t\to t_c$ limit.}
\item{This formula is not trivial, as we are not taking the limit
    $t\to i\infty$ where of course  $\del_t^3 \mathcal{F}_0|_{\rm
      weak}$ reproduces $\kappa_{ttt}$. To remind the reader, here we propose
    a procedure that is valid in the M-theory limit, i.e. strong     string coupling regime.}
\item{Clearly, to evaluate the limit one must know the form of the prepotential $\mathcal{F}_0$ close to the conifold point. In practice, this is most conveniently obtained on the mirror side by solving the Picard-Fuchs equations in a local coordinate of the conifold patch of the mirror CY and picking a symplectic basis of periods by analytic continuation from the large complex structure (LCS) patch. We quickly recall how this is done later. Finally, one can then rewrite it in the Kähler modulus by inverting the mirror map $t(u)$. How all these computations are carried out in detail is described in the original work \cite{Blumenhagen:2025zgf}.}
\end{itemize}

Although in this proceeding  article we focus on the Quintic, in \cite{Blumenhagen:2025zgf} the regularization procedure was generalized to CY manifolds with $h^{1,1}=2$. In those cases, which codimension-two singular point the limit was taken to was crucial in determining whether the regularization procedure worked. Although this issue remains unresolved, we focus here on two simpler questions that were left open.

\section{Regularization scheme for the prepotential \texorpdfstring{${\cal F}_0$}{F₀}}
\label{regScheme}
Since inverting the mirror map is technically challenging, in
\cite{Blumenhagen:2025zgf} the regularization was performed directly
on the mirror side by subtracting the divergent terms in the complex
structure modulus $u$. One may wonder whether, upon reexpressing the result in terms of the Kähler modulus $t$, some of the divergences subtracted in \eqref{regyukawa} could give rise to a finite constant contribution. Since we are interested in the original CY, if that was case the minimal subtraction performed in $u$ would not be sufficient and the regularization
prescription would have to be adjusted. 

\subsection{Limit in complex structure  moduli space}

To show that this does not
occur we revisit the simple example of the quintic and
recall how the computation was carried out in the complex
structure coordinate.
Around the conifold point, $\psi=1$, it is convenient to introduce the
local coordinate $u=1-\psi^{-5}$. Solving the Picard-Fuchs equations
and analytically continuing the symplectic basis to the conifold patch
gives the mirror map \cite{Blumenhagen:2025zgf}
\begin{equation}
  \label{mirrormap}
    t=\frac{X_1}{X_0}=t_c\sum_{m=0}^\infty (u \log u)^m c_m(u)\,,
\end{equation}
where $t_c\approx i1.20812$ and $c_m(u)=\sum_{n=0}^\infty c_{(m,n)} u^n$ denote infinite series in
$u$. Some low-order coefficients are listed in Table~\ref{coeffsmirrormap}.

\begin{table}[ht]
\centering
\begin{tabular}{c|cccc}
  $m \backslash n$ & $n=0$ & $n=1$ & $n=2$ & $n=3$ \\ \hline
  $m=0$ & $1$ & $0.139626$ & $0.074073$ & $0.049758$ \\
  $m=1$ & $-0.0528989$ & $-0.045636$ & $-0.0396285$ & $-0.0351159$ 
\end{tabular}
\caption{Coefficients $c_{(m,n)}$ of the mirror map.}
\label{coeffsmirrormap}
\end{table}

\noindent

Next, recall that the Yukawa coupling can be expressed as \cite{Candelas:1990rm}
\begin{equation}
    \del_t^3 \mathcal{F}_0 |_{\rm weak}=\frac{1}{\omega_0^2}\kappa_{\psi\psi\psi}\frac{1}{(dt/d\psi)^3}
\end{equation}
with
\begin{equation}
    \kappa_{\psi\psi\psi}=\left(\frac{2\pi i}{5}\right)^3\frac{5\psi^2}{1-\psi^5}\,.
\end{equation}
Using $u=1-\psi^{-5}$ and \eqref{mirrormap}, each factor admits an
expansion in powers of $u$ and inverse powers of $\log u$. Specifically,
\eq{\label{expressionsQuintic}
    \frac{dt}{du} &=\log u\left(\frac{d_{-1}(u)}{\log u}+\sum_{n=0}^\infty (u\log u)^n d_n(u)\right)
    \\ \frac{1}{\omega_0^2} &=\sum_{n=0}^\infty (u\log u)^n e_n(u)
    \\ \frac{\kappa_{\psi\psi\psi}}{(\del_\psi u)^3} &= -\frac{1}{25}\frac{1}{u(1-u)^3}\,,
}
where again the $d_n(u)$ and $e_n(u)$ are infinite series in
$u$. Importantly, we will later use the fact that
\begin{equation}\label{d-1exp}
    d_{-1}(u)=\frac{d_{(-1,1)}}{u}+d_{(-1,0)}+\mathcal{O}(u)\,.
\end{equation}
Putting everything together and performing an expansion around $u \to 0$, we obtain
\begin{equation}
    \del_t^3 \mathcal{F}_0 |_{\rm weak}=\frac{1}{u\log^3 u}\left(\sum_{n,k=0}^\infty \frac{(u\log u)^n}{\log^k u}a_{(n,k)}(u)\right) \ ,
\end{equation}
where $a_{(n,k)}(u)$ denotes an infinite series in $u$. The terms split into a
divergent piece when $u\rightarrow 0$, a regular piece which vanishes
as $u\rightarrow0$ and possibly a constant piece. The proposed
regularized Yukawa coupling \eqref{regyukawa} gives the correct result
if, after subtracting the entire divergent piece, there is no constant
contribution surviving in the limit $u\rightarrow0$. This is the case for the quintic, where
\begin{equation}\label{expansionyukawa}
    \partial_t^3 \mathcal{F}_0 \big|_{\rm weak}
    = \sum_{k=0}^{\infty} \frac{a_{(0,k,0)}}{u\,\log^{k+3}u} + {\rm Reg}\,,
\end{equation}
such that the divergent contributions are removed by minimal subtraction. Thus,
\begin{equation}
    Y_{ttt}^{(0)}=-\lim_{t\rightarrow t_c}\left[(\del_t^3 \mathcal{F}_0|_{\rm weak}-\kappa_{ttt})-{\rm Div}\right]=\kappa_{ttt}\,.
\end{equation}
where the term $\kappa_{ttt}$ is subtracted explicitly.

\subsection{Limit in K\"ahler moduli space}

To see what happens when these divergences are rewritten in terms of
$t$,  note that the leading singular behavior in
\eqref{expansionyukawa} reduces to
\begin{equation}
  \label{prepotexp}
    \del_t^3 \mathcal{F}_0 |_{\rm weak}(u)=\frac{1}{u\log^3 u}\left(a_{(0,0,0)}+a_{(0,1,0)}\frac{1}{\log u}+\mathcal{O}\big(\log^{-2} u\big)\right)\,,
\end{equation}
with $a_{(0,0,0)}\approx 67.3587$ and $a_{(0,1,0)}\approx331.301$. In the following, we show that no constant term arises when expressing the expansion \eqref{prepotexp} in terms of the Kähler modulus $t$. Technically, due to the logarithms this computation is highly involved and we proceed by matching the divergences order by
order.

\begin{figure}[ht]
    \centering
    \includegraphics[width=0.7\linewidth]{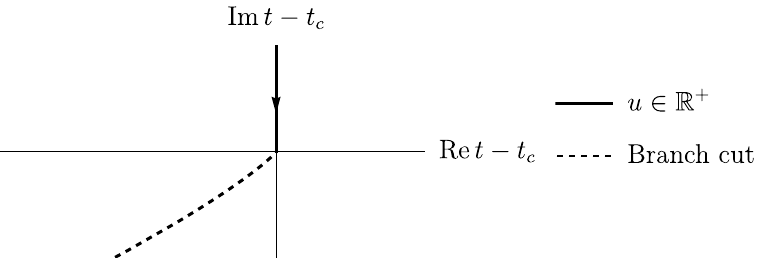}
    \caption{Trajectory $u\to0$ along the real axis and the branch-cut in the $u$-plane, given by the negative real axis, mapped to the $t$-plane using the mirror map \eqref{mirrormap}.}
    \label{fig_branchcut}
\end{figure}

To avoid the complications that arise with complex-valued logarithms, we restrict ourselves to a
particular path on the moduli space in which $u$ goes to zero along
the positive real axis. In terms of the Kähler modulus this trajectory
gets mapped to the positive imaginary axis in the $t$-plane, namely
$t=i|t|$. Furthermore, after mapping the branch-cut in the $u$-plane
(which we choose to be the standard negative real axis) to the
$t$-plane, one can check that the trajectory we have chosen does not
cross the branch cut. This is displayed in Figure \ref{fig_branchcut}.

First, we  determine the leading order expansion for \eqref{prepotexp} in
terms of $|t|$. For small $u$, the mirror map \eqref{mirrormap} takes
the form
\eq{
  \label{deltatexpansion}
    \Delta t:=|t-t_c|=|t_c|\, c_{(1,0)}\, u \log
    u\left(1+\frac{c_{(0,1)}}{c_{(1,0)}}\frac{1}{\log u}+
      \mathcal{O}\big(\log^{-2} u\big)\right)\,,
}
with the values of $c$ given in Table \ref{coeffsmirrormap}. It follows that
\begin{equation}
  \label{logtexp}
    \log \Delta t=\log
    u+L+\mathcal{O}\big(\log^{-2} u\big)\,,\qquad\quad  L\coloneq \log|\log u|+\log |t_c\,c_{(1,0)}|+\frac{c_{(0,1)}}{c_{(1,0)}}\frac{1}{\log u}\,.
\end{equation}
Even though $\log u<0$ and $c_{(1,0)}<0$, eq.\eqref{prepotexp} should be a
real function so it suffices to take the absolute value inside the
logarithms. To match the leading divergence in the limit $u\rightarrow 0$, we evaluate
\eq{
\frac{|t_c|\,c_{(1,0)}\,a_{(0,0,0)}}{\Delta t \log^2 \Delta t}=\frac{1}{ u\log^3 u}& \left[a_{(0,0,0)}-2a_{(0,0,0)}\frac{\log|\log u|}{\log u}\right.
\\&\left.-\left(2\log |t_c
    \,c_{(1,0)}|+\frac{c_{(0,1)}}{c_{(1,0)}}\right)a_{(0,0,0)}\frac{1}{\log
    u}
  +\mathcal{O}\big(\log^{-2} u\big)\right]\,,
}
which indeed reproduces the leading $1/(u\log^3u)$ behavior. However, it also generates two subleading contributions, namely $\log |\log u|/\log u$ and $1/\log u$. The latter is compensated  by including an additional term of the form
\begin{equation}
 \frac{  \alpha \ |t_c| \,c_{(1,0)}}{\Delta t\log^3 \Delta t}=\frac{\alpha}{u\log^3 u}\left(\frac{1}{\log u}+\mathcal{O}\big(\log^{-2} u\big)\right)\,
  \end{equation}
  on the right hand side. Taking into account eq.\eqref{prepotexp},
  by comparison the coefficient $\alpha$ can be determined as
\begin{equation}
    \alpha=a_{(0,1,0)}+\left(2\log |t_c\,c_{(1,0)}|+\frac{c_{(0,1)}}{c_{(1,0)}}\right)a_{(0,0,0)}\,.
\end{equation}
Finally, we need to cancel the $\log|\log u|$ term. From
eq. \eqref{logtexp} it follows that this can be achieved from 
\begin{equation}
    \log|\log \Delta t|=\log|\log u|+L\frac{1}{\log u}+\mathcal{O}\big(\log^{-2} u\big)\,.
\end{equation}
Indeed, one checks that adding
\begin{equation}
    2|t_c|\, c_{(1,0)} \, a_{(0,0,0)}\frac{\log|\log \Delta t|}{\Delta t\log^3 \Delta t}=\frac{1}{u\log^3 u}\left(2a_{(0,0,0)}\frac{\log|\log u|}{\log u}+\mathcal{O}\big(\log^{-2} u\big) \right)\,
\end{equation}
exactly cancels the corresponding contribution.

Hence, we have shown that  up to order $1/(u\log^5 u)$, the expansion \eqref{prepotexp}
can be equivalently expressed in terms of $\Delta t$ as
\begin{equation}
  \label{prepotint}
    \del_t^3 \mathcal{F}_0|_{\rm weak}(\Delta t)=\alpha_1\frac{1}{\Delta t\log^2\Delta t}+\alpha_2\frac{\log|\log \Delta t|}{\Delta t\log^3  \Delta t}+\alpha_3\frac{1}{\Delta t\log^3 \Delta t}+\mathcal{O}\left(\frac{1}{\Delta t\log^4 \Delta t}\right)\,,
\end{equation}
with numerical values
\begin{equation}
    \alpha_1=-4.3048\dots \,,\quad\alpha_2=-8.6096\dots \,,\quad\alpha_3=13.8684\dots\,.
  \end{equation}
 It is important to notice that 
 in the limit   $u\rightarrow 0$, the terms we have introduced in
$\Delta t$ are subleading with respect to the divergences already
matched. Specifically, we have the hierarchy of terms
\begin{equation}
    \frac{1}{\Delta t\log^2\Delta t}\gg\frac{\log|\log \Delta t|}{\Delta t\log^3 \Delta t}\gg\frac{1}{\Delta t\log^3 \Delta t}\gg\dots\,,
\end{equation}
as $\Delta t$ becomes small. Thus, adding the subleading terms
necessary to cancel the $\log|\log u|$ contribution and match the
$1/\log u$ piece does not alter the leading divergent behavior that
had already been fixed.  If one wanted to find the expansion of \eqref{prepotexp} up to
order $1/(u \log^5u)$, the same strategy would apply. Higher orders in
all the expansions would need to be added, together with the
subleading terms $\log|\log \Delta t|/(\Delta t\log^4 \Delta t)$ and
$1/(\Delta t\log^4 \Delta t)$ which would not affect the previous orders. 

\begin{figure}[ht]
    \centering
    \includegraphics[width=0.5\linewidth]{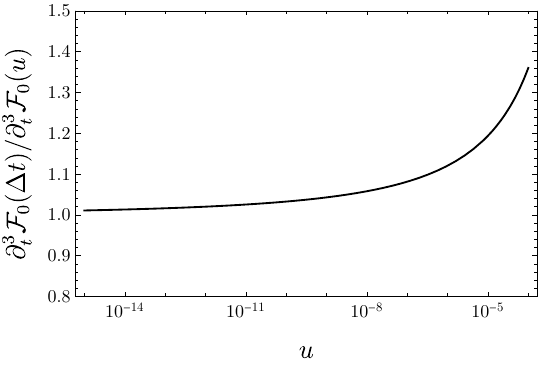}
    \caption{Ratio of eq.\eqref{prepotint} and eq.\eqref{prepotexp} for $u\rightarrow0$.}
    \label{ratioprepotential}
  \end{figure}

In Figure \ref{ratioprepotential} we  depict  the ratio of the
leading divergent terms in \eqref{prepotexp} and the leading expansion
\eqref{prepotint} in $\Delta t$ for small values of $u$ (resp. $\Delta
t$) to confirm that indeed both expressions are equivalent. 
Thus, we have shown  that indeed no finite constant term is generated in this
matching procedure. 
In other words, performing minimal subtraction on
the mirror side does not miss  any constant contribution in the
original Kähler modulus expansion, leaving us with the correct
regularized Yukawa coupling.

We find this quite remarkable, since the structure of the mirror map \eqref{mirrormap} and prepotential \eqref{prepotexp} around the conifold point is highly non-trivial. We believe this further supports our proposed regularization procedure.

\section{Regularization scheme for the 1-loop prepotential \texorpdfstring{${\cal F}_1$}{F₁}}

Another open question is whether this regularization procedure can be extended to higher-genus contributions to the prepotential. As we have seen in eq.\eqref{F1}, similarly to the triple intersection numbers in
$\mathcal{F}_0$, the genus-one prepotential $\mathcal{F}_1$ contains a
classical linear term at weak coupling.
The  question is whether this term can also be obtained
via a suitable regularization from the GV Schwinger integral \eqref{F1_Schwinger}.

This integral is divergent, so one has to introduce a regulator, $\epsilon$. Then, for the integral over $s$, we obtain
\begin{equation}
    \mathcal{F}_1=\sum_{(\beta,n)\neq (0,0)}\left(\frac{1}{12}\alpha_0^\beta+\alpha_1^\beta\right)\left(\gamma_E+\log(-\epsilon Z_{\beta,n})+\mathcal{O}(\epsilon)\right)\,,
\end{equation}
with $\gamma_E$ the Euler-Mascheroni constant. There are two different
contributions, one coming from $D2$-$D0$-brane bound states and one from pure $D0$-branes. An analogous analysis has been performed in
\cite{Blumenhagen:2023tev} for the non-compact case of the resolved
conifold, the only difference being that now one
has to perform the sum over the GV invariants so that
the two different contributions read
\eq{
\mathcal{F}_1^{D0,D2}
&= \sum_{\beta \neq 0} \left( \frac{1}{12}\alpha_0^\beta + \alpha_1^\beta \right)
\left( -\pi i\, \beta t + \log\big(1 - e^{2\pi i \beta t}\big) \right) \,,
\\[0.1cm]
\mathcal{F}_1^{D0}
&= \frac{1}{12}\log \mu \, .
}
 where $\mu=g_s/( e^{\gamma_E}\epsilon)$.  Note that due to the logarithmic nature of the integral,
the regulator appears explicitly when
trying to define a regularization procedure for $\mathcal{F}_1$. This did not occur for $\mathcal{F}_0$, where all the terms
involving the regulator $\epsilon$ vanish upon $\zeta$-function
regularization. As discussed in \cite{vandeHeisteeg:2023dlw}, operators such as $R^4$ in eight-dimensional maximal supergravity or $R^2$ in four dimensions with $N=2$ receive logarithmic corrections from massless states. In order to properly account for this divergence, we effectively incorporate a moduli dependent regulator $\epsilon(t)$.

Then, we can get information about the linear term in \eqref{F1} by taking a
derivative with respect to $t$, yielding a function that is analogous
to the Yukawa coupling so that we define
\eq{
    C_t&:=\del_t\left(\mathcal{F}_1^{D2,D0}+\mathcal{F}_1^{D0}\right)\\[0.3em]
    &= - 2\pi i \sum_{\beta\neq 0}\left(\frac{1}{12}\alpha_0^\beta+\alpha_1^\beta\right)\beta\left(\frac{1}{2}+\frac{e^{2\pi i \beta t}}{1-e^{2\pi i\beta t}}\right)- \frac{1}{12}\del_t\log\epsilon(t)\,.
  }
  For the simple choice  $\alpha_0=1$, $\alpha_1=0$,
  similarly to the case of $Y_{ttt}$, we recover
the formal second Chern class  of the resolved conifold \cite{Gopakumar:1998ki}, $c_2=-1$.
Since the expression already contains the correct world-sheet
instanton contributions, Emergence suggests that 
after regularization,  the zero point  contribution
\begin{equation}
    C_t^{(0)}:=- \pi i \sum_{\beta\neq 0}\left(\frac{1}{12}\alpha_0^\beta+\alpha_1^\beta\right)\beta\,\bigg|_{\rm Reg}- \ \frac{1}{12}\del_t\log\epsilon(t)\,
  \end{equation}
  is equal to   $(2\pi i) c_2/24$. 
Carrying out the analogous  heuristic computation as in the last section motivates the 
  regularization
  \begin{equation}\label{c2reg}
    C_t^{(0)}=-(2\pi i)\lim_{t\to
    t_c}\left[\left(\left. \frac{1}{2\pi i} \del_t\mathcal{F}_1\right|_{\rm
    weak}-\frac{c_2}{24}\right)-{\rm Div}+\frac{1}{24\pi i}\del_t\log\epsilon(t)\right]\,.
\end{equation}

Similarly to the case of the genus zero prepotential, this computation is to be done in the strongly coupled M-theory limit, where we need to integrate out the full tower of light states. Of course, the tools to carry this computation are not yet available, but we can use the result from
the weakly coupled regime using mirror symmetry. The main difference with respect
to $\mathcal{F}_0$ is that due to the
logarithmic divergence of the initial Schwinger integral, now there is some
freedom in choosing the regulator.

\subsection{Limit in complex structure  moduli space}

Let us again focus on the Quintic, where we can rely on the
work of \cite{Bershadsky:1993ta, Hosono:1994ax} to compute $\mathcal{F}_1$\footnote{For higher genus computations and related work see e.g. \cite{Bershadsky:1993cx,Katz:1999xq,Yamaguchi:2004bt}.}  in
terms of the period data of the CY that we already have
available. Let $K(t,\bar{t})$ be the Kähler potential for the metric
$G_{t\bar{t}}=\del_t\bar{\del}_{\bar{t}}K(t,\bar{t})$, which can be
computed from the symplectic basis of periods in the standard way
\begin{equation}
    e^{-K}=-i \Pi^\dagger\,\Sigma\,\Pi\,,\qquad\qquad \Sigma=\begin{pmatrix}
        0 & 0 &0 & 1 \\
        0 & 0 &1 & 0 \\
        0 & -1 &0 & 0 \\
        -1 & 0 &0 & 0 \\
    \end{pmatrix}\,.
\end{equation}
Then, $\mathcal{F}_1$ was determined from a topological string
amplitude computation in \cite{Bershadsky:1993ta} as
\begin{equation}
    \mathcal{F}_1=\pi i \log\left[e^{\left(3+h_{1,1}-\frac{\chi}{12}\right)K}\det (G^{-1})|f|^2\right]\,,
\end{equation}
where $f$ is a holomorphic function, the holomorphic anomaly, which should be determined by imposing the appropriate behavior at the singular points of the moduli space. Focusing now on the mirror quintic, the LCS behavior $\mathcal{F}_1\rightarrow 2 \pi i \ \frac{50}{24}(t+\bar{t})$, together with imposing regularity at the orbifold point $\psi=0$, fixes $f$ to be
\begin{equation}
    f=\psi^{\frac{62}{3}}(1-\psi^{-5})^{-\frac{1}{6}}\,.
\end{equation}
In the topological limit $\bar{t}\rightarrow -i\infty$, the first
derivative of the genus one prepotential reads
\begin{equation}
    \del_t\mathcal{F}_1= 2 \pi i \  \frac{\del \psi}{\del t}\del_\psi\log\left[e^{\frac{31}{3}K}\det \left(G_{\psi\bar{\psi}}^{-1}\right)^{\frac{1}{2}}\psi^{\frac{62}{6}}(1-\psi^{-5})^{-\frac{1}{12}}\right]\,.
\end{equation}
This expression holds everywhere in moduli space, which allows us to
verify our regularization procedure when approaching the
conifold. Working again in the coordinate $u$, for small values we find the expansion
\begin{equation}
  \label{deltF1expinu}
    \left.\del_t\mathcal{F}_1\right|_{\rm weak}=\frac{2 \pi i }{u\log u}\left(\sum_{n,k=0}^\infty \frac{(u\log u)^n}{\log^k u}\tilde a_{(n,k)}(u)\right)\,,
\end{equation}
where as usual $\tilde a_{(n,k)}(u)$ denotes a polynomial in $u$. For
$n=0$, all the constant terms in $\tilde a_{(0,k)}(u)$ lead to singular
terms which we minimally subtract. Particularly interesting  is
the term with $n=1$, $k=0$, as it leads to a potential constant
contribution, concretely to
\begin{equation}
\label{aconifold}
    \tilde a_{(1,0)}(u)=-0.0219561+\mathcal{O}(u)\,.
\end{equation}

At first sight, such a  constant contribution seems to spoil our proposed
regularization. However, just for ${\cal F}_1$ there is
this extra contribution from the  choice of regulator $\epsilon(t)$,
which we can
choose appropriately in order to cancel the constant term that
appears in \eqref{c2reg}.
To make this concrete, in hindsight we  choose the regulator to be of
the form $\epsilon(t) = \Delta t\, \epsilon$, where
$\Delta t$ is defined as in \eqref{deltatexpansion}\footnote{Note that
  in the limit $t\to t_c$ this regulator is well defined
  since $\Delta t=|t-t_c|>0$.}.
Then, the regulator dependent term in \eqref{c2reg} reads
\begin{equation}
  \label{reginu}
    \lim_{t\to t_c}\frac{1}{24\pi i}\del_t\log\epsilon(t)\approx\lim_{u\to0}\frac{1}{24\pi i}\frac{du}{dt}\del_u\left[\log u+\log\!|\!\log u|+\log |t_c\,c_{(1,0)}|+\frac{c_{(0,1)}}{c_{(1,0)}}\frac{1}{\log u}\right]\,,
\end{equation}
where we have plugged the expansion \eqref{logtexp} at leading
order. We take the derivative with respect to $u$ and note that
inverting the Jacobian in \eqref{expressionsQuintic} leads to the series expansion
\begin{equation}
    \frac{du}{dt}=d_{1}(u)+\sum_{n=2}^\infty(u\log u)^n d_n(u)\,,
\end{equation}
where the $d_n$ are some infinite series in $u$ with $d_1(u)\approx
i1.65545 u+\mathcal{O}(u^2)$ being the inverse expansion of \eqref{d-1exp}. Finally, plugging everything in
\eqref{reginu} yields
\begin{equation}
    \lim_{t\to t_c}\frac{1}{24\pi i}\del_t\log\epsilon(t)\approx\lim_{u\to0}\frac{1}{24\pi i}(i1.65545+\mathrm{Reg.})\approx 0.0219561\,.
\end{equation}
Moreover, it is a straightforward computation to convince oneself that no new divergences are
introduced in \eqref{c2reg} by choosing this regulator. 

In summary, combining the expansion \eqref{deltF1expinu} with this
regulator in \eqref{c2reg} leads to a cancellation of the constant
term and therefore yields precisely $c_2^{(0)}=(2\pi i) c_2/24$ after
performing minimal subtraction. As in other cases with such
logarithmic divergences, due to inevitable  regulator dependence 
one looses the predictivity of the formalism.
Thus, the main observation is that precisely
for the case of ${\cal F}_1$, where a constant term
could emerge, the regulator appears and helps canceling it. 

\subsection{Limit in K\"ahler moduli space}
Finally, we can do a similar analysis to that of Section
\ref{regScheme} and check whether performing the minimal subtraction
in terms of $u$ instead of the original Kähler modulus $t$ misses
any possible constant contribution. The aim is to find the leading
divergence expansion of \eqref{deltF1expinu} in terms of $t$. We
content ourselves to work up to $1/(u\log^2u)$-order, i.e.
\begin{equation}
  \label{divexpF1}
    \del_t \mathcal{F}_1\bigr|_{\rm weak}(u)=\frac{2\pi i }{u\log u}\left(\tilde a_{(0,0,0)}+\tilde a_{(0,1,0)}\frac{1}{\log u}+\mathcal{O}\big(\log^{-2} u\big)\right)\,,
\end{equation}
with $\tilde a_{(0,0,0)}=-0.207529$ and $\tilde a_{(0,1,0)}=-0.340241$. We
proceed in an analogous manner to the case of
$\mathcal{F}_0$. Recalling the expansion \eqref{deltatexpansion}, it is clear that
\begin{equation}
    \frac{|t_c|\, c_{(1,0)}\, \tilde a_{(0,0,0)}}{\Delta t}=\frac{1}{u\log u}\left(\tilde a_{(0,0,0)} -\tilde a_{(0,0,0)}\frac{c_{(0,1)}}{c_{(1,0)}}\frac{1}{\log u}+\mathcal{O}\big(\log^{-2} u\big)\right)\,
\end{equation}
matches the leading divergence. In order to match the $1/(u\log^2 u)$
divergence,  we expand $\Delta t\log \Delta t$ and find
\begin{equation}
    \frac{|t_c|c_{(1,0)}\,\beta}{\Delta t\log \Delta t}=\frac{\beta}{u\log u}\left(\frac{1}{\log u}+\mathcal{O}\big(\log^{-2} u\big)\right)\,.
\end{equation}
By setting
\begin{equation}
    \beta=\tilde a_{(0,1,0)}+\tilde a_{(0,0,0)}\frac{c_{(0,1)}}{c_{(1,0)}}\,,
\end{equation}
the expansion \eqref{divexpF1} can equivalently be  expressed in terms of $t$ as
\begin{equation}
  \label{F1int}
    \del_t \mathcal{F}_1(\Delta t)=\alpha_1\frac{2 \pi i }{\Delta t}+\alpha_2\frac{2 \pi i }{\Delta t\log\Delta t}+\mathcal{O}\left(\frac{1}{\Delta t\log^2\Delta t}\right)
\end{equation}
with $\alpha_1=|t_c|c_{(1,0)}\, \tilde a_{(0,0,0)}\approx 0.0132$ and
$\alpha_2=|t_c|c_{1,0}\,\beta\approx -0.0132$. Note that this is in
accordance with the expected behavior of $\mathcal{F}_1$ around the conifold, namely
\begin{equation}
\,\mathcal{F}_1\sim-\frac{1}{12}\log(t-t_c)\,,
\end{equation}
as e.g. found  in \cite{Huang:2006hq}.

\begin{figure}[ht]
    \centering
    \includegraphics[width=0.5\linewidth]{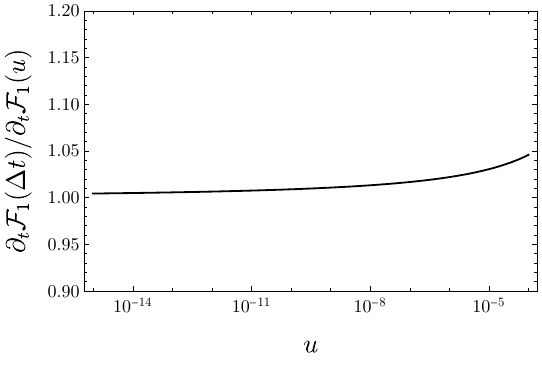}
    \caption{Ratio of eq.\eqref{F1int} and eq.\eqref{divexpF1} for $u\rightarrow0$.}
    \label{ratioF1}
\end{figure}

In Figure \ref{ratioF1} we show the ratio of both expansions to
confirm that around $u\to 0$ both expressions are equivalent.
Again,
no finite constant appears when inverting the mirror map or,
differently said, performing the minimal subtraction on the mirror
side, which is arguably easier, does not miss  any possible
constant contribution in the original Kähler expansion. Once more, we remark that the fact that the minimal subtraction does not depend on whether we consider the Kähler or the mirror complex structure moduli space provides non-trivial support for our proposed regularization of the sum over the GV invariants.

\section{Conclusion}
Motivated by questions raised during  talks given on the
original paper \cite{Blumenhagen:2025zgf},
in this article we have extended the emergence computation
to the linear term in the 1-loop prepotential
${\cal F}_1$. In contrast to ${\cal F}_0$, here the result
is less predictive as the regulator  plays an important role.
However, it is still not in contradiction with the proposed regularization, as precisely for the case where a potential
constant could appear, the appearing regulator can cancel it. We have shown this in detail for
the Quintic threefold and  expect this computation to be
analogous for the other $14$ CY threefolds with $h^{1,1}=1$.
In addition, for both ${\cal F}_0$ and ${\cal F}_1$, we have shown that performing the limit of the regularization procedure in the more tractable complex structure moduli space of the mirror CY ultimately yields the same result as taking the limit in the physical K\"ahler moduli space. This a postereori confirms the method used in \cite{Blumenhagen:2025zgf}.

It would also be interesting to gain a better understanding of the regularization procedure for CY manifolds with multiple K\"ahler moduli.
The methods to deal with such cases were
introduced in \cite{Blumenhagen:2025zgf} and
two CYs with $h_{11}=2$ were discussed in detail,
with the final results being encouraging but not yet fully conclusive. That's because new issues, like the choice of the codimension two
singular point in the K\"ahler moduli space and the
correct path towards, which remain to be explored in more detail. We hope to return to these in the future.

Nevertheless, we think it is fair to say that by now we have
promising evidence for the validity of the
M-theoretic Emergence Proposal, at least for $1/2$-BPS saturated amplitudes. For the general non-BPS amplitudes it is expected that one really needs a fully fledged  QG theory of  M-theory. From this broader perspective, the emerging nature of $1/2$-BPS amplitudes is just a reflection of the emerging nature of interactions in quantum M-theory. The expectation is that, analogous to the BFSS matrix theory, an ordinary commuting space-time emerges as a particular solution to the equations of motion of such a theory. Interactions, including gravity, then arise from quantum effects, potentially through the evaluation of appropriate Schwinger integrals over the  spectrum of quantum M-theory.

\paragraph{Acknowledgments.}
We are grateful to Thorsten Schimannek  and
Niccol\`o Cribiori for discussions on the emergence of the
one-loop prepotential.  We also thank Antonia Paraskevopoulou for helpful comments on the draft. 
R.B. acknowledges the hospitality of the Corfu Summer Institute 2025, where part of
the research presented here has been initiated.


\newpage

\begin{thebibliography}{99}



%\cite{Palti:2019pca}
\bibitem{Palti:2019pca}
E.~Palti,
``The Swampland: Introduction and Review,''
Fortsch. Phys. \textbf{67}, no.6, 1900037 (2019)
%doi:10.1002/prop.201900037
[arXiv:1903.06239 [hep-th]].
% 844 citations counted in INSPIRE as of 24 Feb 2024

%\cite{vanBeest:2021lhn}
\bibitem{vanBeest:2021lhn}
M.~van Beest, J.~Calder\'on-Infante, D.~Mirfendereski and I.~Valenzuela,
``Lectures on the Swampland Program in String Compactifications,''
Phys. Rept. \textbf{989}, 1-50 (2022)
%doi:10.1016/j.physrep.2022.09.002
[arXiv:2102.01111 [hep-th]].
% 286 citations counted in INSPIRE as of 24 Feb 2024

%\cite{Agmon:2022thq}
\bibitem{Agmon:2022thq}
N.~B.~Agmon, A.~Bedroya, M.~J.~Kang and C.~Vafa,
``Lectures on the string landscape and the Swampland,''
[arXiv:2212.06187 [hep-th]].
% 81 citations counted in INSPIRE as of 24 Feb 2024



% \cite{Ooguri:2006in}
\bibitem{Ooguri:2006in}
H.~Ooguri and C.~Vafa,
``On the Geometry of the String Landscape and the Swampland,''
Nucl. Phys. B \textbf{766}, 21-33 (2007)
%doi:10.1016/j.nuclphysb.2006.10.033
[arXiv:hep-th/0605264 [hep-th]].
% 960 citations counted in INSPIRE as of 24 Feb 2024


%\cite{Lee:2019wij}
\bibitem{Lee:2019wij}
S.~J.~Lee, W.~Lerche and T.~Weigand,
``Emergent strings from infinite distance limits,''
JHEP \textbf{02}, 190 (2022)
%doi:10.1007/JHEP02(2022)190
[arXiv:1910.01135 [hep-th]].
%174 citations counted in INSPIRE as of 24 Feb 2024

%\cite{Blumenhagen:2023tev}
\bibitem{Blumenhagen:2023tev}
R.~Blumenhagen, N.~Cribiori, A.~Gligovic and A.~Paraskevopoulou,
``Demystifying the Emergence Proposal,''
JHEP \textbf{04}, 053 (2024)
%doi:10.1007/JHEP04(2024)053
[arXiv:2309.11551 [hep-th]].


%\cite{Blumenhagen:2023xmk}
\bibitem{Blumenhagen:2023xmk}
R.~Blumenhagen, N.~Cribiori, A.~Gligovic and A.~Paraskevopoulou,
``Emergent M-theory limit,''
Phys. Rev. D \textbf{109}, no.2, L021901 (2024)
%doi:10.1103/PhysRevD.109.L021901
[arXiv:2309.11554 [hep-th]].

%\cite{Dvali:2007hz}
\bibitem{Dvali:2007hz}
G.~Dvali,
``Black Holes and Large N Species Solution to the Hierarchy Problem,''
Fortsch. Phys. \textbf{58}, 528-536 (2010)
%doi:10.1002/prop.201000009
[arXiv:0706.2050 [hep-th]].
% 430 citations counted in INSPIRE as of 24 Feb 2024

  %\cite{Dvali:2007wp}
\bibitem{Dvali:2007wp}
G.~Dvali and M.~Redi,
``Black Hole Bound on the Number of Species and Quantum Gravity at LHC,''
Phys. Rev. D \textbf{77}, 045027 (2008)
%doi:10.1103/PhysRevD.77.045027
[arXiv:0710.4344 [hep-th]].
% 273 citations counted in INSPIRE as of 24 Feb 2024

%\cite{Heidenreich:2017sim}
\bibitem{Heidenreich:2017sim}
B.~Heidenreich, M.~Reece and T.~Rudelius,
``The Weak Gravity Conjecture and Emergence from an Ultraviolet Cutoff,''
Eur. Phys. J. C \textbf{78}, no.4, 337 (2018)
%doi:10.1140/epjc/s10052-018-5811-3
[arXiv:1712.01868 [hep-th]].
% 131 citations counted in INSPIRE as of 24 Feb 2024

%\cite{Grimm:2018ohb}
\bibitem{Grimm:2018ohb}
T.~W.~Grimm, E.~Palti and I.~Valenzuela,
``Infinite Distances in Field Space and Massless Towers of States,''
JHEP \textbf{08}, 143 (2018)
%doi:10.1007/JHEP08(2018)143
[arXiv:1802.08264 [hep-th]].
%269 citations counted in INSPIRE as of 24 Feb 2024

%\cite{Heidenreich:2018kpg}
\bibitem{Heidenreich:2018kpg}
B.~Heidenreich, M.~Reece and T.~Rudelius,
``Emergence of Weak Coupling at Large Distance in Quantum Gravity,''
Phys. Rev. Lett. \textbf{121}, no.5, 051601 (2018)
%doi:10.1103/PhysRevLett.121.051601
[arXiv:1802.08698 [hep-th]].
%152 citations counted in INSPIRE as of 24 Feb 2024


%\cite{Marchesano:2022axe}
\bibitem{Marchesano:2022axe}
F.~Marchesano and L.~Melotti,
``EFT strings and emergence,''
JHEP \textbf{02}, 112 (2023)
%doi:10.1007/JHEP02(2023)112
[arXiv:2211.01409 [hep-th]].
% 17 citations counted in INSPIRE as of 24 Feb 2024

%\cite{Castellano:2022bvr}
\bibitem{Castellano:2022bvr}
A.~Castellano, A.~Herr\'aez and L.~E.~Ib\'a\~nez,
``The emergence proposal in quantum gravity and the species scale,''
JHEP \textbf{06}, 047 (2023)
%doi:10.1007/JHEP06(2023)047
[arXiv:2212.03908 [hep-th]].
% 28 citations counted in INSPIRE as of 24 Feb 2024



 %\cite{Blumenhagen:2023yws}
\bibitem{Blumenhagen:2023yws}
R.~Blumenhagen, A.~Gligovic and A.~Paraskevopoulou,
``The emergence proposal and the emergent string,''
JHEP \textbf{10}, 145 (2023)
%doi:10.1007/JHEP10(2023)145
[arXiv:2305.10490 [hep-th]].
% 13 citations counted in INSPIRE as of 24 Feb 2024


%\cite{Hattab:2023moj}
\bibitem{Hattab:2023moj}
J.~Hattab and E.~Palti,
``On the particle picture of Emergence,''
JHEP \textbf{03}, 065 (2024)
%doi:10.1007/JHEP03(2024)065
[arXiv:2312.15440 [hep-th]].
% 19 citations counted in INSPIRE as of 20 Feb 2026

%\cite{Hattab:2024thi}
\bibitem{Hattab:2024thi}
J.~Hattab and E.~Palti,
``Emergence in string theory and Fermi gases,''
JHEP \textbf{07}, 144 (2024)
%doi:10.1007/JHEP07(2024)144
[arXiv:2404.05176 [hep-th]].
%12 citations counted in INSPIRE as of 20 Feb 2026


%\cite{Hattab:2024chf}
\bibitem{Hattab:2024chf}
J.~Hattab and E.~Palti,
``Emergent potentials and non-perturbative open topological strings,''
JHEP \textbf{10}, 195 (2024)
%doi:10.1007/JHEP10(2024)195
[arXiv:2408.12302 [hep-th]].
%10 citations counted in INSPIRE as of 20 Feb 2026

%\cite{Hattab:2024ssg}
\bibitem{Hattab:2024ssg}
J.~Hattab and E.~Palti,
``Notes on integrating out M2 branes,''
Eur. Phys. J. C \textbf{85}, no.1, 107 (2025)
%doi:10.1140/epjc/s10052-025-13827-5
[arXiv:2410.15809 [hep-th]].
%7 citations counted in INSPIRE as of 16 Mar 2026

%\cite{vanMuiden:2026nsp}
\bibitem{vanMuiden:2026nsp}
J.~van Muiden,
``Quantum M2-branes and Holography,''
[arXiv:2603.14544 [hep-th]].
% 0 citations counted in INSPIRE as of 19 Mar 2026


%\cite{Blumenhagen:2024lmo}
\bibitem{Blumenhagen:2024lmo}
R.~Blumenhagen, N.~Cribiori, A.~Gligovic and A.~Paraskevopoulou,
``Reflections on an M-theoretic Emergence Proposal,''
PoS \textbf{CORFU2023}, 238 (2024)
%doi:10.22323/1.463.0238
[arXiv:2404.05801 [hep-th]].
% 13 citations counted in INSPIRE as of 18 Feb 2026




%\cite{Banks:1996vh}
\bibitem{Banks:1996vh}
T.~Banks, W.~Fischler, S.~H.~Shenker and L.~Susskind,
``M theory as a matrix model: A Conjecture,''
Phys. Rev. D \textbf{55}, 5112-5128 (1997)
%doi:10.1103/PhysRevD.55.5112
[arXiv:hep-th/9610043 [hep-th]].
% 3132 citations counted in INSPIRE as of 24 Feb 2024

%\cite{Taylor:2001vb}
\bibitem{Taylor:2001vb}
W.~Taylor,
``M(atrix) Theory: Matrix Quantum Mechanics as a Fundamental Theory,''
Rev. Mod. Phys. \textbf{73}, 419-462 (2001)
%doi:10.1103/RevModPhys.73.419
[arXiv:hep-th/0101126 [hep-th]].
% 330 citations counted in INSPIRE as of 24 Feb 2024



%\cite{Green:1997as}
\bibitem{Green:1997as}
M.~B.~Green, M.~Gutperle and P.~Vanhove,
``One loop in eleven-dimensions,''
Phys. Lett. B \textbf{409}, 177-184 (1997)
%doi:10.1016/S0370-2693(97)00931-3
[arXiv:hep-th/9706175 [hep-th]].
% 334 citations counted in INSPIRE as of 24 Feb 2024




%\cite{Blumenhagen:2024ydy}
\bibitem{Blumenhagen:2024ydy}
R.~Blumenhagen, N.~Cribiori, A.~Gligovic and A.~Paraskevopoulou,
``Emergence of R$^{4}$-terms in M-theory,''
JHEP \textbf{07}, 018 (2024)
%doi:10.1007/JHEP07(2024)018
[arXiv:2404.01371 [hep-th]].
% 22 citations counted in INSPIRE as of 18 Feb 2026



%\cite{Artime:2026new}
\bibitem{Artime:2026new}
M.~Artime, R.~Blumenhagen and P.~Leivadaros, work in progress.
%Eur. Phys. J. C \textbf{85}, no.7, 730 (2025)
%doi:10.1140/epjc/s10052-025-14452-y
%[arXiv:2504.05392 [hep-th]].
% 3 citations counted in INSPIRE as of 18 Feb 2026

%\cite{Etheredge:2024tok}
\bibitem{Etheredge:2024tok}
M.~Etheredge, B.~Heidenreich, T.~Rudelius, I.~Ruiz and I.~Valenzuela,
``Taxonomy of infinite distance limits,''
JHEP \textbf{03}, 213 (2025)
%doi:10.1007/JHEP03(2025)213
[arXiv:2405.20332 [hep-th]].
% 30 citations counted in INSPIRE as of 20 Feb 2026


%\cite{Kiritsis:2000zi}
\bibitem{Kiritsis:2000zi}
E.~Kiritsis, N.~A.~Obers and B.~Pioline,
``Heterotic / type II triality and instantons on K(3),''
JHEP \textbf{01}, 029 (2000)
%doi:10.1088/1126-6708/2000/01/029
[arXiv:hep-th/0001083 [hep-th]].
%75 citations counted in INSPIRE as of 20 Feb 2026

%\cite{Artime:2025egu}
\bibitem{Artime:2025egu}
M.~Artime, R.~Blumenhagen and A.~Paraskevopoulou,
``Emergence of $F^4$-couplings in heterotic/type IIA dual string theories,''
Eur. Phys. J. C \textbf{85}, no.7, 730 (2025)
%doi:10.1140/epjc/s10052-025-14452-y
[arXiv:2504.05392 [hep-th]].
%3 citations counted in INSPIRE as of 18 Feb 2026



%\cite{Gopakumar:1998ii}
\bibitem{Gopakumar:1998ii}
R.~Gopakumar and C.~Vafa,
``M theory and topological strings. 1.,''
[arXiv:hep-th/9809187 [hep-th]].
% 541 citations counted in INSPIRE as of 24 Feb 2024

%\cite{Gopakumar:1998jq}
\bibitem{Gopakumar:1998jq}
R.~Gopakumar and C.~Vafa,
``M theory and topological strings. 2.,''
[arXiv:hep-th/9812127 [hep-th]].
% 644 citations counted in INSPIRE as of 24 Feb 2024

%\cite{Blumenhagen:2025zgf}
\bibitem{Blumenhagen:2025zgf}
R.~Blumenhagen and A.~Gligovic,
``Emergence of CY triple intersection numbers in M-theory,''
JHEP \textbf{10}, 048 (2025)
%doi:10.1007/JHEP10(2025)048
[arXiv:2506.20725 [hep-th]].
% 3 citations counted in INSPIRE as of 18 Feb 2026

 %\cite{Gopakumar:1998ki}
\bibitem{Gopakumar:1998ki}
R.~Gopakumar and C.~Vafa,
``On the gauge theory / geometry correspondence,''
Adv. Theor. Math. Phys. \textbf{3}, 1415-1443 (1999)
%doi:10.4310/ATMP.1999.v3.n5.a5
[arXiv:hep-th/9811131 [hep-th]].
%741 citations counted in INSPIRE as of 18 Feb 2026

\bibitem{Candelas:1990rm}
P.~Candelas, X.~C.~De La Ossa, P.~S.~Green and L.~Parkes,
``A Pair of Calabi-Yau manifolds as an exactly soluble superconformal theory,''
Nucl. Phys. B \textbf{359}, 21-74 (1991)
%doi:10.1016/0550-3213(91)90292-6
%1114 citations counted in INSPIRE as of 18 Feb 2026

%\cite{Morrison:1994fr}
\bibitem{Morrison:1994fr}
D.~R.~Morrison and M.~R.~Plesser,
``Summing the instantons: Quantum cohomology and mirror symmetry in toric varieties,''
Nucl. Phys. B \textbf{440}, 279-354 (1995)
%doi:10.1016/0550-3213(95)00061-V
[arXiv:hep-th/9412236 [hep-th]].
%291 citations counted in INSPIRE as of 16 Mar 2026

%\cite{vandeHeisteeg:2023dlw}
\bibitem{vandeHeisteeg:2023dlw}
D.~van de Heisteeg, C.~Vafa, M.~Wiesner and D.~H.~Wu,
``Species scale in diverse dimensions,''
JHEP \textbf{05}, 112 (2024)
%doi:10.1007/JHEP05(2024)112
[arXiv:2310.07213 [hep-th]].
%77 citations counted in INSPIRE as of 16 Mar 2026

%\cite{Bershadsky:1993ta}
\bibitem{Bershadsky:1993ta}
M.~Bershadsky, S.~Cecotti, H.~Ooguri and C.~Vafa,
``Holomorphic anomalies in topological field theories,''
Nucl. Phys. B \textbf{405}, 279-304 (1993)
%doi:10.1016/0550-3213(93)90548-4
[arXiv:hep-th/9302103 [hep-th]].
% 527 citations counted in INSPIRE as of 20 Feb 2026

%\cite{Hosono:1994ax}
\bibitem{Hosono:1994ax}
S.~Hosono, A.~Klemm, S.~Theisen and S.~T.~Yau,
``Mirror symmetry, mirror map and applications to complete intersection Calabi-Yau spaces,''
Nucl. Phys. B \textbf{433}, 501-554 (1995)
%doi:10.1016/0550-3213(94)00440-P
[arXiv:hep-th/9406055 [hep-th]].
%344 citations counted in INSPIRE as of 05 Mar 2026

%\cite{Bershadsky:1993cx}
\bibitem{Bershadsky:1993cx}
M.~Bershadsky, S.~Cecotti, H.~Ooguri and C.~Vafa,
``Kodaira-Spencer theory of gravity and exact results for quantum string amplitudes,''
Commun. Math. Phys. \textbf{165}, 311-428 (1994)
%doi:10.1007/BF02099774
[arXiv:hep-th/9309140 [hep-th]].
%1177 citations counted in INSPIRE as of 05 Mar 2026

%\cite{Katz:1999xq}
\bibitem{Katz:1999xq}
S.~H.~Katz, A.~Klemm and C.~Vafa,
``M theory, topological strings and spinning black holes,''
Adv. Theor. Math. Phys. \textbf{3}, 1445-1537 (1999)
%doi:10.4310/ATMP.1999.v3.n5.a6
[arXiv:hep-th/9910181 [hep-th]].
%242 citations counted in INSPIRE as of 05 Mar 2026

%\cite{Yamaguchi:2004bt}
\bibitem{Yamaguchi:2004bt}
S.~Yamaguchi and S.~T.~Yau,
``Topological string partition functions as polynomials,''
JHEP \textbf{07}, 047 (2004)
%doi:10.1088/1126-6708/2004/07/047
[arXiv:hep-th/0406078 [hep-th]].
%125 citations counted in INSPIRE as of 05 Mar 2026

%\cite{Huang:2006hq}
\bibitem{Huang:2006hq}
M.~x.~Huang, A.~Klemm and S.~Quackenbush,
``Topological string theory on compact Calabi-Yau: Modularity and boundary conditions,''
Lect. Notes Phys. \textbf{757}, 45-102 (2009)
%doi:10.1007/978-3-540-68030-7{\_}3
[arXiv:hep-th/0612125 [hep-th]].
%179 citations counted in INSPIRE as of 20 Feb 2026

\end{thebibliography}
\end{document}